# High-speed measurement of rifle primer blast waves

Michael W. Courtney, Ph.D., U.S. Air Force Academy, 2354 Fairchild Drive, USAF Academy, CO, 80840
Michael.Courtney@usafa.edu

Amy C. Courtney, Ph.D., Force Protection Industries, Inc., 9801 Highway 78, Ladson, SC 29456
amy_courtney@post.harvard.edu

**Abstract:** This article describes a method and results for direct high-speed measurements of rifle primer blast waves employing a high-speed pressure transducer located at the muzzle to record the blast pressure wave produced by primer ignition. Our key findings are: 1) Most of the primer models tested show 5-12% standard deviation in the magnitudes of their peak pressure. 2) For most primer types tested, peak pressure magnitudes are well correlated with measured primer masses so that significant reductions in standard deviation are expected to result from sorting primers by mass. 3) A range of peak pressures from below 200 psi to above 500 psi is available in different primer types.

**Keywords**: *rifle primer, blast pressure, primer strength, muzzle velocity variations*

## I. Introduction

Over the years various surrogates have been used to quantify and compare performance of rifle primers including measuring velocity and standard deviation when the primer alone propelled a projectile from a gun barrel,(1) measuring velocity, pressure, and standard deviation produced by a given primer in combination with a given powder charge and bullet,(2)(3) and measuring the size of the visible primer flash in photographs.(2)(3) This article presents a method and results for direct high-speed measurements of rifle primer blast waves employing a high-speed pressure transducer located at the muzzle to record the blast pressure wave produced by primer detonation and by showing that mass sorting produces a smaller deviation in peak primer pressures.

It is commonly reported that choosing the least powerful primer that can reliably ignite a powder charge often produces the smallest standard deviations in muzzle velocity, thus the smallest vertical dispersions at long range. Two causal hypotheses have emerged for this observation. Lapua's published brochure on the .308 Winchester Palma Case featuring a small rifle primer pocket describes the idea that small rifle primers themselves simply exhibit less variations. The other hypothesis is (in the words of German Salazar), "accuracy is more easily found when the influence of the primer on the overall pressure of the load is minimized."(3) The data presented here is inconclusive regarding which hypothesis is more correct; however, the measurement method presented could be used, together with mass sorting and measurement of velocity standard deviations to determine which hypothesis is better supported in a given cartridge and load.

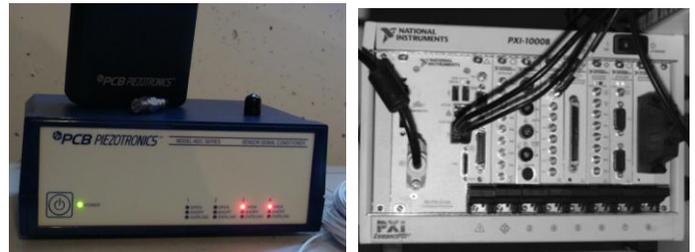

*Figure 1: Left: High-speed pressure transducer (on top) and signal conditioning unit. Right: Fast waveform digitizer in PXI system.*

## II. Method

Rifle primers work by the impact detonation of high-explosive compounds (usually a combination of lead styphnate and lead azide in modern primers), which then ignites the propellant charge. The measurement method is simple: a firearm loaded with a primed cartridge case without any gunpowder or projectile has all the essential elements of an explosive driven shock tube whose shock wave is emitted from the muzzle after the primer is detonated by the firing pin. The blast wave measured at the muzzle depends on the strength of the primer without the confounding factors (bore friction, neck tension, powder charge, bullet bearing surface, cartridge case variations, etc.) that affect other methods of inferring primer strength and consistency.

Here, a Remington 700 ADL chambered in .308 Winchester with a 22" barrel is used for the test platform.  Tests on large rifle primers employ R-P brass with the pockets uniformed with the Redding tool, and the flash hole deburred with a handheld center drill from the outside and an oversized drill bit from the inside.  Primers are loaded into the case with an RCBS Rockchucker reloading press.  Tests on small rifle primers employ the Lapua Palma case featuring a small rifle primer pocket prepared in the same manner.  Reported masses are determined on an AccuLab VIC-123 with a resolution of 1 mg.

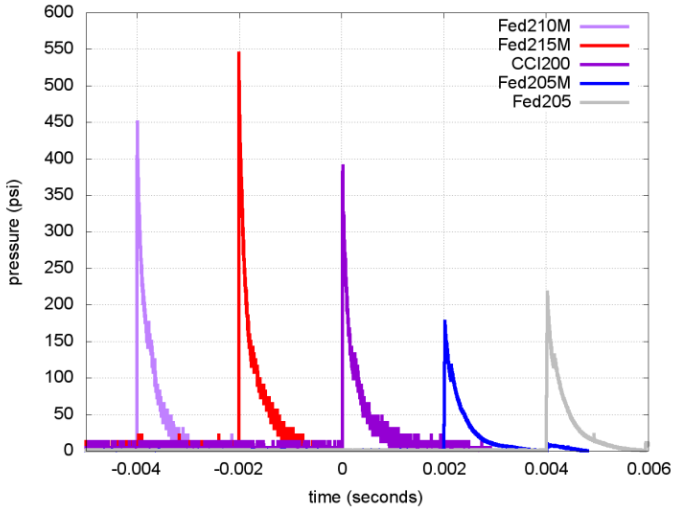

*Figure 2: Typical blast pressure waveforms measured for several rifle primer types.  The detonation times have been shifted in 0.002 second (2 millisecond) increments to better visualize and compare waveform shapes.*

The blast pressure measurements presented here result from using high-speed pressure transducers (PCB 102B and PCB 102B15) specifically designed for measuring the very fast pressure transients associated with explosive detonations and other shock waves.  The pressure transducer is placed coaxially with the rifle barrel and directly facing the muzzle with no separation between the end of the barrel and pressure transducer.  A cable connects the transducer to a signal conditioning unit (PCB 842C) which produces a calibrated voltage output which is then digitized with a National Instruments PXI-5105 fast analog to digital converter operating at a rate of 1 million samples per second.  The voltage waveform is saved as a file for later conversion to pressure using the calibration certificate provided by the manufacturer with each pressure sensor.  A high-speed pressure transducer, signal conditioning unit, and fast waveform digitizer are shown in Figure 1.

### III. Results

Figure 2 shows blast pressure waveforms for several rifle primer types.  The waveforms are combined on a single graph to facilitate comparison.  Dozens of these waveforms were measured for the study reported here, but rather than show all the graphs, it is more revealing to characterize the waveform shapes with their key parameters and then discuss the average and standard deviation because these best characterize primer strength and consistency.

Simple blast waves are usually characterized by peak overpressure, duration, and impulse (the area under the curve of pressure vs. time).  Since the durations and basic shapes are all nearly the same for all the pressure waveforms, the impulse is nearly proportional to the peak pressure, and the peak pressure is the main distinguishing characteristic of the blast wave.  Therefore, we will focus on the average peak magnitude and the standard deviation of peak magnitudes for each primer type.

| Primer | Peak Pressure (psi) | SD (psi) | SD (%) |
|---|---|---|---|
| **Fed210M** | 421.7 | 32.4 | 7.7% |
| **Fed215M** | 552.7 | 27.8 | 5.0% |
| **CCI200** | 371.4 | 39.2 | 10.7% |
| **CCI250** | 520.3 | 58.6 | 11.3% |
| **Fed205M (lot 1CW306)** | 176.5 | 20.4 | 11.6% |
| **Fed205** | 213.0 | 15.0 | 7.1% |
| **CCI450** | 232.3 | 15.1 | 6.5% |
| **Fed205M (lot 13X416)** | 208.0 | 15.0 | 7.2% |
| **Rem 7 ½** | 334.0 | 27.0 | 8.1% |

*Table 1: Peak pressure averages and standard deviations from the mean (SD) with a sample size of 10.*

Standard deviation is a statistically valid measure of the variability of a quantity.  (Extreme spread is a popular metric of a quantity's variability, but lacks statistical validity.)  Our ability to focus on peak blast pressure alone might be a consequence of all the primers having a composition dominated by lead styphnate and lead azide.  Future work with lead-free primers might reveal different waveform shapes or different durations, in which case the impulse might need careful consideration as well.

Table 1 shows average peak pressures along with standard deviations from the mean for a selection of both large and small rifle primers. As expected, large rifle primers produce stronger blast waves than small primers, and "magnum" rifle primers (Fed215M, CCI250, CCI450) produce stronger blast waves than non-magnum primers of the same size. There are significant differences in the standard deviations observed for different primer types, and it is notable that so-called "Match" primers are not always more consistent than non-match primers. It is also notable that the two different lots of Fed205M primers show more than a 15% difference in their means.

To determine whether the variation in peak blast pressure is caused by variations in primer mass, we measured the mass of each primer, plotted peak blast pressure vs. primer mass, and performed a linear regression to determine the resulting slope (increase in pressure for every milligram increase in mass) and the correlation coefficient, R. This analysis was performed for every primer type reported here, and a representative graph for Fed210M primers is shown in Figure 3.

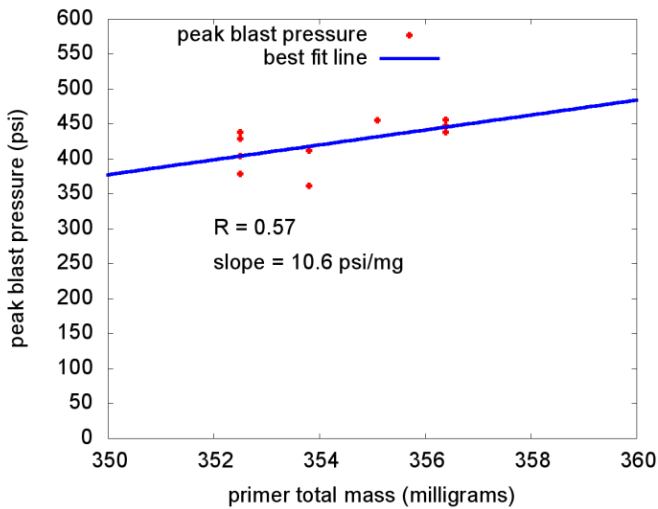

*Figure 3: Peak pressure vs. primer mass for Fed 210M: correlation implies reduction in variation with mass sorting of primers.*

The hypothesis that the variations in mass of the primer are mainly due to variations in mass of the priming compound can be tested by measuring the primers after firing and estimating the mass of priming compound as the mass lost during primer detonation. For the Fed210M primer, the total primer mass is strongly correlated with the mass of priming compound with R = 0.92. Of course, the peak pressure is expected to be more strongly correlated with actual mass of the priming compound than with the total primer mass, as shown in Figure 4. However, this is of limited practical use, since the actual explosive mass cannot be easily determined until after the primer is detonated.

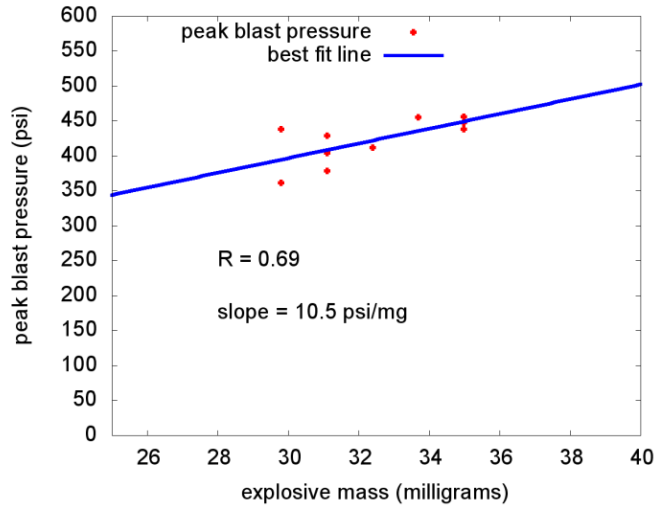

*Figure 4: Peak pressure vs. mass of explosive compound (as determined by mass loss in detonation) for Federal 210M. This graph shows the variation in mass of the priming compound as well as the resulting variation in peak blast pressure.*

These results suggest that sorting the Federal 210M primer by mass will reduce variations in peak blast pressure. To test this idea, we sorted 100 Fed210M primers to obtain 10 measuring within a 1 mg range centered at 355 mg. These primers exhibited an average peak blast pressure of 440.7 psi and a standard deviation of 18.3 psi. Mass sorting to within 1 mg reduced the SD significantly from 32.4 psi to 18.3 psi. Since the slope of peak pressure vs. mass is 10.5 psi/mg, the best we would expect is a variation of 10.5 psi for a group of mass sorted primers with 1 mg variation if peak pressure variations were due to mass variations alone; however, variations in other factors increase the pressure variations somewhat. It seems there is even more room to reduce peak pressure variations by sorting by mass even more finely, but this would require a more precise scale.

High-speed measurement of rifle primer blast wavesTable 2 shows that, of the primers tested, peak blast pressure is well-correlated with the total primer mass in all cases, except for the CCI 450 primer. This is highly suggestive that sorting primers by mass will usually be effective at reducing variations in peak primer blast pressure. Mass sorting will likely be more effective at reducing peak primer blast variations for primers with a higher correlation coefficient, and primers with a larger dependence of pressure on primer mass (slope).

| Primer | Total Mass (mg) | SD % | R | Slope (psi/mg) |
|---|---|---|---|---|
| Fed210M | 354.2 | 0.49% | 0.57 | 10.6 |
| Fed215M | 363.3 | 0.61% | 0.84 | 9.3 |
| CCI200 | 371.4 | 0.51% | 0.81 | 18.0 |
| CCI250 | 337.5 | 0.60% | 0.65 | 18.8 |
| Fed205M (lot 1CW306) | 236.3 | 0.72% | 0.76 | 9.1 |
| Fed205 | 239.5 | 0.30% | 0.66 | 14.0 |
| CCI450 | 239.3 | 0.56% | 0.03 | N/A |
| Fed205M (lot 13X416) | 234.9 | 0.47% | 0.59 | 8.0 |
| Rem 7 ½ | 229.0 | 0.94% | 0.69 | 8.5 |

*Table 2: Correlations and slopes between peak pressure and primer total mass. Note that small primers (Fed 205M, Fed205, and CCI450) are over 100 mg lighter than large primers in total mass; however, the actual mass of the explosive priming compound is 33-40 mg in large rifle primers and 12-16 mg in small rifle primers. The remaining mass is the metal in the primer cup and anvil.*

### IV. Discussion

Key findings are: 1) Primer types tested show 5-12% standard deviation in the magnitudes of their peak pressure. 2) For most primer types, peak pressure magnitudes are well correlated with measured primer masses so that significant reductions in standard deviation are expected to result from sorting primers by mass at 1 mg or smaller increments. 3) A range of peak pressures from below 200 psi to above 500 psi is available in different primer types.

Since we purchased 100 samples of Lapua .308 Win Palma brass to test small primers, some comments on this new offering from Lapua are warranted. In *The Rifleman's Journal*,(4) German Salazar reports the weight variation to be very uniform for a sample of 24 cases obtained from Lester Bruno (Bruno Shooter's Supply). He reports that all the weights are between 173.7 and 174.6 grains, which he considers satisfactory without sorting. In our 100 samples, the lightest case was 170.6 grains and the heaviest case was 172.6 grains with an average weight of 171.4 and a standard deviation of 0.43 grains. In contrast, a sample of 100 cases of regular Lapua .308 Win brass (large rifle primer pockets) showed an average mass of 171.6 grains, a standard deviation of 0.21 grains, and an extreme spread of only 0.9 grains. We are unsure why our sample of the Palma brass differs so much from Salazar's both in average weight and in variation, or why the Laupa's regular .308 Win brass shows less than ½ the variation as their Palma offering.

One always wants more data, and we expect very few will read this article without wishing to see results for some of the primers they use regularly or wishing that we had presented results for larger sample sizes. Instead of presenting an exhaustive sampling of available primers, we have presented a method of measuring the magnitude and variation of primer performance more quantitatively and independently of confounding factors than prior methods.

It is clear that for many primers, sorting by mass would have the effect of reducing variations in primer strength; however, the degree of reduction this would have on muzzle velocity variation is an open question. We believe significant reductions in muzzle velocity variation are possible, but this opinion is more based on our own passion for precision and desire for uniformity in every component than on conclusive data which is not yet available.

However, it should be pointed out that since most of the slopes of peak primer blast pressure vs. primer mass are between 8 and 20 psi/mg, achieving variations in peak primer blast pressure below 10 psi will likely require sorting primers into groups on the order of ¼ to ½ a milligram. (The expected variation in peak primer blast pressure will be no smaller than roughly the variation in primer mass times the slope of the regression line for pressure vs. mass.) Confidently sorting into groups with mass ranges this small likely requires a scale with a precision of 0.1 milligram. This precision requires use of an analytical laboratory scale, and scales with this precision start at about $1000.

With mass sorting on a sufficiently fine scale, the experiment to determine whether primer strength or uniformity has the bigger effect on muzzle velocity variations is simple: prepare a group of test samples with both unsorted and sorted primers and compare their standard deviations from the mean muzzle velocities. If one obtains the same result for numerous loads, then confidence builds that the result will likely be the same for other loads as well. In contrast, it is also possible that some loads are more sensitive to primer uniformity than others, in which case it might not be not possible to know whether mass sorting of primers will benefit a specific load until it is tried for that load.

High-speed pressure sensors and digitizers aren't cheap, but for parties interested in repeating and extending this work, progress is possible for much less than the tens of thousands of dollars of equipment used here. For well under five thousand dollars, one could configure a workable system with a USB-based digitizer with a sample rate of at least 1 million samples per second such as the National Instruments USB-5132, a single high speed transducer such as the PCB 102A04, and a single channel signal conditioning unit. Direct measurement of primer blast waves is not necessary to determine whether mass sorting on a fine scale would benefit a given load, but it should prove useful in more quickly determining the strength and uniformity of new primer introductions.

Primer science seems to go in cycles where it takes several decades to empirically determine the most accurate primer for a given application and then a new generation of less accurate primers is developed to meet new demands (non-mercuric, non-corrosive). (5) Having determined much about selecting the best primer for a given application from the current crop of lead-based primers, shooters may soon find ourselves repeating the process with lead-free offerings. Hopefully, contributions to understanding the basic scientific principles of primer performance will hasten the process of primer improvement and selection compared with the much longer trial and error process of past generations.

**Acknowledgements:**
The authors acknowledge and appreciate the use of test equipment from Force Protection Industries, Inc. We are also grateful to German Salazar for providing helpful suggestions and comments on the manuscript and to Leo Ahearn of the Colorado Rifle Club for providing resources on short notice.

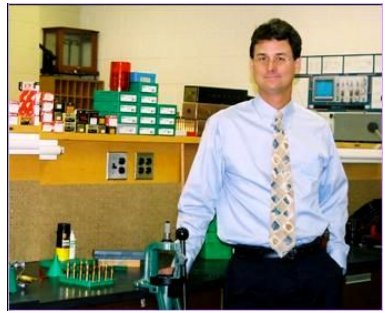

*About the Authors:*
*Michael Courtney* is shown in his ballistics lab. He earned a PhD in experimental physics from MIT and published papers in theoretical astrophysics, experimental atomic physics, chaos theory, quantum theory, acoustics, ballistics, traumatic brain injury, and education. He served as director of the Forensic Science Program at Western Carolina University and as a college professor, teaching physics, calculus, statistics, and forensic science. He currently serves at the United States Air Force Academy where he teaches Mathematics and maintains an active research program in ballistics and blast injury.

*Amy Courtney* describes our ballistics research at the United States Military Academy where she served as a physics professor. She earned a MS in medical engineering from Harvard University and a PhD in medical engineering and medical physics in a joint Harvard/MIT program. Amy has also served as a research scientist at the Cleveland Clinic and currently serves as a scientific consultant for Force Protection Industries, Inc.

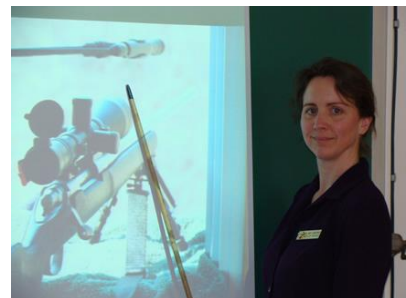